%% file: main.tex
\documentclass[conference]{IEEEtran}

\usepackage{cite}
\usepackage{url}
\usepackage[bookmarks=false]{hyperref}
\usepackage{balance}
\usepackage{xspace}

\usepackage{subcaption}
\usepackage{graphicx}
\usepackage{booktabs}
\usepackage{multirow}
\usepackage{threeparttable}
\usepackage[table]{xcolor}

\usepackage{algorithm}
\usepackage{algorithmicx}
\usepackage{algpseudocode}
\makeatletter
\newcommand\fs@betterruled{%
  \def\@fs@cfont{\bfseries}\let\@fs@capt\floatc@ruled
  \def\@fs@pre{\vspace*{5pt}\hrule height.8pt depth0pt \kern2pt}%
  \def\@fs@post{\kern2pt\hrule\relax}%
  \def\@fs@mid{\kern2pt\hrule\kern2pt}%
  \let\@fs@iftopcapt\iftrue}
\floatstyle{betterruled}
\restylefloat{algorithm}
\makeatother

\usepackage{tcolorbox,tikz}
\usepackage{pgfplots}
\usepackage{pgfplotstable}
\usepackage{filecontents}
\usepgfplotslibrary{statistics}
\pgfplotsset{compat=1.14} 
\usepackage{silence}
\usepackage{ifthen}

\usepackage{tcolorbox}
\tcbuselibrary{skins}
\newenvironment{summarybox}
{\begin{tcolorbox}
[enhanced,arc=0mm,colback=gray!10,frame hidden,overlay unbroken={%
    \draw[thick,black] (interior.north west)--(interior.south west);
},left=2pt,right=0pt,top=0pt,bottom=0pt,before={\vspace{3pt}\noindent},after={\vspace{0pt}}]}
{\end{tcolorbox}}

\newenvironment{summary}
{\vspace{5pt}\noindent\begin{summarybox} Summary:}
{\end{summarybox}\vspace{-5pt}}

\newcommand\numberToBeChecked[1]{\textcolor{black}{#1}}

\newcommand{\nbSeedPapers}{\numberToBeChecked{13}\xspace}
\newcommand{\nbPapersACM}{\numberToBeChecked{1,926}\xspace}
\newcommand{\nbPapersIEEE}{\numberToBeChecked{729}\xspace}
\newcommand{\nbPapersScopus}{\numberToBeChecked{1,909}\xspace}
\newcommand{\nbPapersFromAllEngines}{\numberToBeChecked{2,843}\xspace}
\newcommand{\nbPapersAfterExclusionCriteria}{\numberToBeChecked{523}\xspace}
\newcommand{\nbPapersAfterInclusionCriteria}{\numberToBeChecked{54}\xspace}
\newcommand{\nbPapersFinalSelection}{\numberToBeChecked{54}\xspace}

\usepackage[inline]{enumitem}

\newboolean{showcomments}
\setboolean{showcomments}{true}

\ifthenelse{\boolean{showcomments}}
  {\newcommand{\nb}[3]{
  {\color{#2}\small\fbox{\bfseries\sffamily\scriptsize#1}}
  {\color{#2}\sffamily\small$\triangleright~$\textit{\small #3}$~\triangleleft$\GenericWarning{}{LaTeX Warning: #1: #3}}
  }
  \newcommand{\todo}[1]{{\color{red}{TODO: #1}}\GenericWarning{}{LaTeX Warning: TODO: #1}}
  }
  {\newcommand{\nb}[3]{}
  \newcommand{\todo}[1]{}
  }

\begin{document}


\title{Evaluating Code Readability and Legibility: \\An Examination of Human-centric Studies}

\author{\IEEEauthorblockN{Delano Oliveira\IEEEauthorrefmark{1}, Reydne Bruno\IEEEauthorrefmark{1}, Fernanda Madeiral\IEEEauthorrefmark{2}, Fernando Castor\IEEEauthorrefmark{1}}
\IEEEauthorblockA{\IEEEauthorrefmark{1}Federal University of Pernambuco, Recife, Brazil, dho@cin.ufpe.br, reydne.bruno@gmail.com, castor@cin.ufpe.br}
\IEEEauthorblockA{\IEEEauthorrefmark{2}KTH Royal Institute of Technology, Stockholm, Sweden, fer.madeiral@gmail.com}
}


\maketitle

\begin{abstract}
Reading code is an essential activity in software maintenance and evolution. Several studies with human subjects have investigated how different factors, such as the employed programming constructs and naming conventions, can impact code readability, i.e., what makes a program easier or harder to read and apprehend by developers, and code legibility, i.e., what influences the ease of identifying elements of a program. These studies evaluate readability and legibility by means of different comprehension tasks and response variables. In this paper, we examine these tasks and variables in studies that compare programming constructs, coding idioms, naming conventions, and formatting guidelines, e.g., recursive vs. iterative code. To that end, we have conducted a systematic literature review where we found \nbPapersFinalSelection relevant papers. Most of these studies evaluate code readability and legibility by measuring the correctness of the subjects' results (\numberToBeChecked{83.3\%}) or simply asking their opinions (\numberToBeChecked{55.6\%}). Some studies (\numberToBeChecked{16.7\%}) rely exclusively on the latter variable. There are still few studies that monitor subjects' physical signs, such as brain activation regions (\numberToBeChecked{5\%}). Moreover, our study shows that some variables are multi-faceted. For instance, correctness can be measured as the ability to predict the output of a program, answer questions about its behavior, or recall parts of it. These results make it clear that different evaluation approaches require different competencies from subjects, e.g., tracing the program vs. summarizing its goal vs. memorizing its text. To assist researchers in the design of new studies and improve our comprehension of existing ones, we model program comprehension as a learning activity by adapting a preexisting learning taxonomy. This adaptation indicates that some competencies, e.g., tracing, are often exercised in these evaluations whereas others, e.g., relating similar code snippets, are rarely targeted.
\end{abstract}


\section{Introduction}


Understanding a program usually requires reading code. A program might be easier or harder to read depending on its  \textit{readability}, i.e., what makes a program easier or harder to read and apprehend by developers, and \textit{legibility}, i.e., what influences the ease of identifying elements of a program. Different factors can influence code readability and legibility, such as which constructs are employed~\cite{Gopstein2017,Ajami2019}, how the code is formatted with whitespaces~\cite{Arab1992,Wang2014,Bauer2019}, and identifier naming conventions~\cite{Teasley1994,Blinman2005,Lawrie2007,Binkley2009,Ceccato2009,Scanniello2013,Avidan2017,Beniamini2017,Schankin2018,Hofmeister2019}.
Researchers have conducted empirical studies to investigate several of those factors, where different but functionally equivalent ways of writing code are compared, e.g., recursive vs. iterative code \cite{Benander1996} and abbreviated vs. word
identifier names \cite{Hofmeister2019}. These studies involve asking subjects to perform one or more tasks related to source code and assessing their understanding or the effort involved in the tasks.

There are systematic literature reviews about program comprehension studies \cite{siegmund2016,schroter2017comprehending}. To the best of our knowledge, however, no previous work has investigated how code readability and legibility are evaluated in studies comparing different ways of writing code, in particular, what kinds of tasks these studies conduct and what response variables they employ. Analyzing these tasks and variables can help researchers identify limitations in the evaluations of previous studies, gauge their comprehensiveness, and improve our understanding of what said tasks and variables evaluate. 

In this paper, we investigate how code readability and legibility are evaluated in studies that aim to identify ways of writing code that are more readable or legible than others. Our goal is to study what types of tasks are performed by human subjects in those studies, what cognitive skills are required from them, and what response variables those studies employ. To achieve this goal, we carried out a systematic literature review, which started with \nbPapersFromAllEngines documents, where \nbPapersFinalSelection papers are relevant for our research---these papers are the primary studies of our study. Our results show that \numberToBeChecked{40} primary studies involve asking subjects to provide information about a program, e.g., to predict its output or the values of its variables, or to provide a high level description of its functioning. In addition, \numberToBeChecked{30} studies involve asking subjects to provide personal opinion, and \numberToBeChecked{15} studies involve asking subjects to act on code. Most studies employ combinations of these types of tasks. 

When considering response variables, the most common approach is to verify if the subjects are able to correctly provide information about the code or act on it. Correctness is the response variable of choice in \numberToBeChecked{45} studies. Moreover, \numberToBeChecked{27} studies measure the time required to perform tasks as an imprecise proxy to the effort required by these tasks. Besides, mirroring the aforementioned tasks where subjects are required to provide their personal opinions, \numberToBeChecked{30} studies employ opinion as a response variable. Additionally, \numberToBeChecked{9} of the \numberToBeChecked{54} studies use opinion as the sole response variable. Furthermore, studies evaluating legibility tend to employ opinion as a response variable proportionally more often than studies focusing on readability. Also, there are still relatively few studies that monitor subjects' physical signs, such as brain activation regions (\numberToBeChecked{5\%}) \cite{Siegmund2017,Yeh2017,Fakhoury2019b}. The analysis of the response variables reveals that they are multi-faceted. For example, correctness can be measured as the ability to predict the output of a program, answer questions about its general behavior, precisely recall specific parts of it, among other things. 

To assist researchers in the design of new studies and improve our comprehension of existing ones, we adapted an existing learning taxonomy to the context of program comprehension. This adapted taxonomy indicates that some competencies, e.g., tracing, are often exercised in the studies whereas others, e.g., relating similar code snippets, are rarely targeted. The taxonomy also indicates that \numberToBeChecked{37\%} of the primary studies have a narrow focus, requiring a single cognitive skill from the subjects, even though program comprehension is a complex activity. Additionally, the taxonomy highlights the tendency of the studies evaluating readability and legibility to focus on comprehension tasks that do not contextualize the information that is apprehended. In software development practice, program comprehension is often associated with other tasks such as reengineering an existing program or extending it. Existing studies rarely go this far.

\section{Code Readability and Legibility}\label{sec:code-readability-and-legibility}

In software engineering, the terms readability, legibility, understandability, and comprehensibility have overlapping meanings. For example, Buse and Weimer \cite{Buse2010} define ``\textit{\textbf{readability} as a human judgment of how easy a text is to understand}''. In a similar vein, Almeida et al. \cite{de2003best} affirm that ``\textit{\textbf{legibility} is fundamental to code maintenance; if source code is written in a complex way, understanding it will require much more effort}''. In addition, Lin and Wu \cite{lin2008} state that 
``\textit{``Software \textbf{understandability}'' determines whether a system can be understood by other individuals easily, or whether artifacts of one system can be easily understood by other individuals}''. Xia et al. \cite{xia2018} treat comprehension and understanding as synonyms, expressing that ``\textit{Program \textbf{comprehension} (aka., program understanding, or source code comprehension) is a process where developers actively acquire knowledge about a software system by exploring and searching software artifacts, and reading relevant source code and/or documentation}''. 

In linguistics, the concept of text comprehension is similar to program comprehension in software engineering. Gough and Tunmer \cite{gough1986decoding} state that ``\textit{comprehension (not reading comprehension, but rather linguistic comprehension) is the process by which given lexical (i.e., word) information, sentences and discourses are interpreted}''. However, Hoover and Gough \cite{hoover1990simple} further elaborate on that definition and claim that ``\textit{decoding and linguistic comprehension are separate components of reading skill}''. This claim highlights the existence of two separate processes during text comprehension: (i) decoding the words/symbols and (ii) interpreting them and sentences formed by them. 
DuBay \cite{dubay2004principles} separates these two processes and defines them as \textbf{legibility}, which concerns typeface, layout, and other aspects related to the identification of elements in text, and \textbf{readability}, that is, what makes some texts easier to read than others. In a similar vein, for Tekif \cite{tekfi1987readability}, legibility studies are mainly concerned with typographic and layout factors while readability studies concentrate on the linguistic factors. 

These two perspectives also apply to programs. We can find both the visual characteristics and linguistic factors in source code, although with inconsistent terminology. 
For example, Daka et al. \cite{daka2015} affirm that ``\textit{the visual appearance of code in general is referred to as its readability}''. The authors clearly refer to legibility (in the design/linguistics sense) but employ the term ``readability'' possibly because it is more often used in the software engineering literature.

Based on the differences between the terms ``readability'' and ``legibility'' that are well-established in other areas such as linguistics \cite{dubay2004principles}, design \cite{strizver2013type}, human-computer interaction \cite{zuffi2007human}, and education \cite{tekfi1987readability}, we believe that the two terms should have clear, distinct, albeit related, meanings also in the area of software engineering. On the one hand, the structural and semantic characteristics of the source code of a program that affect the ability of developers to understand it while reading the code, e.g., programming constructs, coding idioms, and meaningful identifiers, impact its 
\textbf{readability}. On the other hand, the visual characteristics of the source code of a program,  which affect the ability of developers to identify the elements of the code while reading it, such as line breaks, spacing, alignment, indentation, blank lines, identifier capitalization, impact its \textbf{legibility}. Hereafter, we employ these two terms according to these informal definitions. 

\section{Methodology}\label{sec:methodology}

Our goal is to investigate how human-centric studies evaluate whether a certain way of writing code is more readable or legible than another functionally equivalent one. More specifically, we investigate what tasks are performed by human subjects and how their performance is evaluated in empirical studies aiming to evaluate readability and legibility. We focus on studies that directly compare two or more different ways of writing code and have a focus on low-level source code elements, without accounting for tools, documentation, or higher level issues (details are further presented in this section). We address two research questions in this paper:

\begin{description}
\item[RQ1] What are the tasks performed by human subjects in empirical studies? 

\item[RQ2] What are the response variables of these studies?
\end{description}

\begin{figure}[t]
    \centering
    \includegraphics[width=0.426\textwidth]{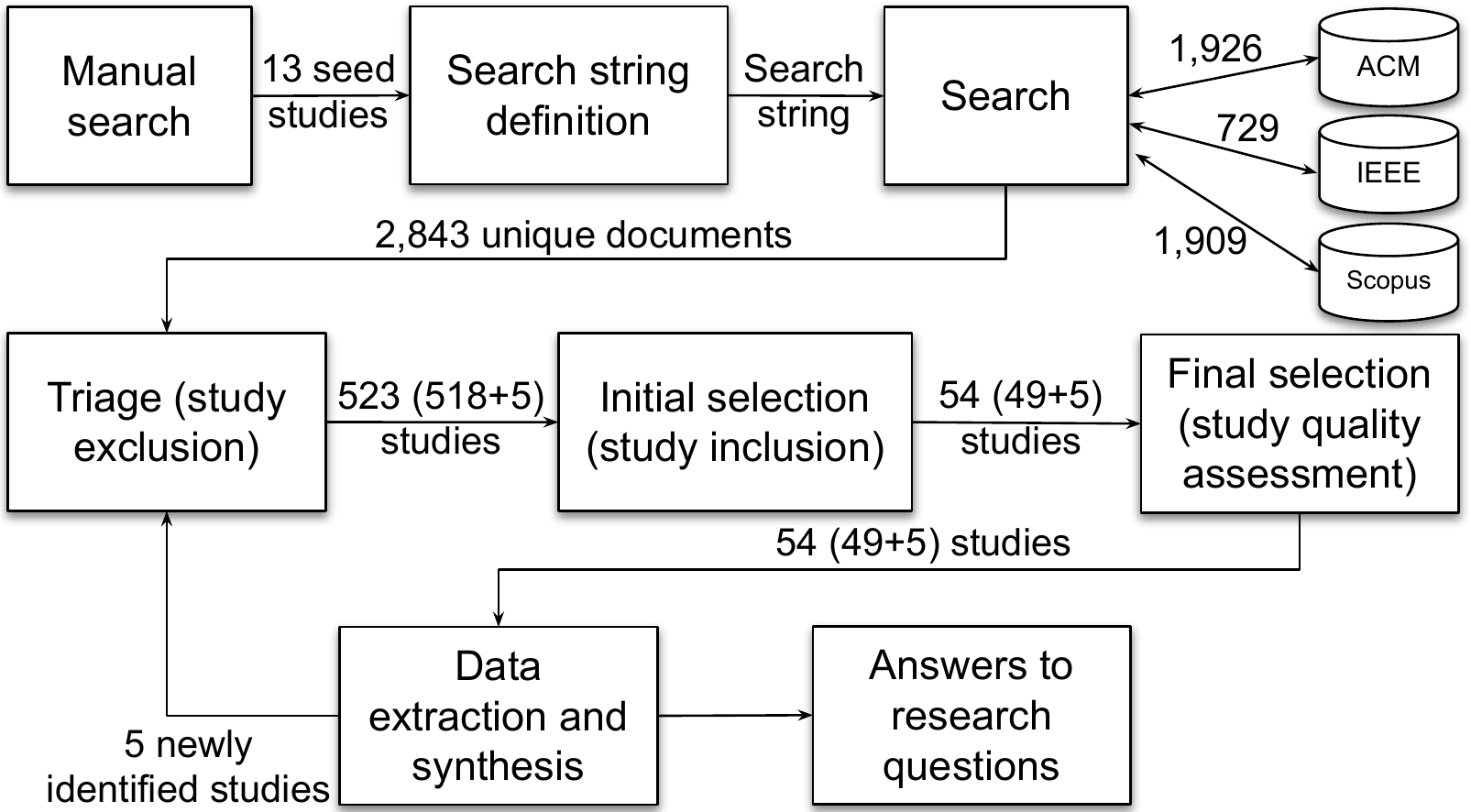}
    \caption{Systematic literature review roadmap.}
    \label{fig:review_roadmap}
\end{figure}

To answer our research questions, we conducted a systematic literature review, which was designed following the guidelines proposed by Kitchenham et al. \cite{Kitchenham2015}. \autoref{fig:review_roadmap} presents the roadmap of our review including all steps we followed.
First, we performed the selection of studies. We started with a manual search for papers to further be used as seed papers, so that a search string could be defined, and automatic search could be performed on search engines (\autoref{sec:search}). We retrieved \nbPapersFromAllEngines documents with the automatic search, which passed through 1) a triage for study exclusion (\autoref{sec:triage}), 2) an initial study selection where inclusion criteria were applied (\autoref{sec:inclusion}), and 3) a final study selection where we evaluated the quality of the studies based on a number of criteria (\autoref{sec:quality-assessment}). Then, the selected \nbPapersFinalSelection papers were analyzed (\autoref{sec:study_analysis}). We extracted data from the papers and synthesized it to answer our research questions. We detail these steps in the following sections.
It is worth mentioning that we did not leverage systematic review tools, like Parsifal \cite{parsifal}, because we were not aware of them at the time.

\subsection{Search Strategy}\label{sec:search}

Our search strategy is composed of three parts: a manual search to gather seed studies, the definition of a generic search string, and the automatic search in search engines.
First, we performed the manual search for seed studies aiming to find terms for the definition of a generic search string. For that, we chose the following top-tier software engineering conferences: ICSE, FSE, MSR, ASE, ISSTA, OOPSLA, ICSME, ICPC, and SANER. Then, the first two authors of this paper looked at the papers published in these conferences in the last three years and a half (from 2016 to June 2019), and selected the ones that would help us answer our research questions. We also included one paper from TSE, which we already knew is relevant to our research. This process resulted in \nbSeedPapers seed papers.

The title and keywords of the seed papers were analyzed, and then we extracted the general terms related to our research questions. We chose general terms as a conservative way to gather as many papers as possible that can fit within the scope of our study; otherwise, we would delimit our set of studies based on specific topics. We used the resulting terms to build the following search string:

\begin{center}
\small

$Title(ANY(terms))~OR~Keywords(ANY(terms))$,

where
$terms = \{$
``code comprehension'',
``code understandability'',

``code understanding'',
``code readability'',

``program comprehension'',
``program understandability'',

``program understanding'',
``program readability'',

``programmer experience'' $\}$
\end{center}

We did not include terms with ``legibility'' in the search string. Most of the papers with this word in the title or keywords are related to linguistics or computational linguistics. In these fields, researchers use this term with a different meaning than what would be expected in a software engineering paper. Using it in our search string would drastically increase the number of false positives. Also, we did not include terms with ``software'', e.g., ``software readability'', because software is broader than source code and program.

Finally, we performed an automatic search for studies using our generic search string adapted for three search engines: ACM Digital Library\cite{acm}, IEEE Explore\cite{ieee}, and Scopus\cite{scopus}. We retrieved \nbPapersACM, \nbPapersIEEE, and \nbPapersScopus documents, respectively, in September 2, 2019. Since a given document might be retrieved from more than one engine, we unified the output of the engines to eliminate duplicates, which resulted in \nbPapersFromAllEngines unique documents. The \nbSeedPapers seed papers were returned by the automatic search.

\subsection{Triage (Study Exclusion)}\label{sec:triage}

The \nbPapersFromAllEngines documents retrieved with our automatic search passed through a triage process so that we could discard clearly irrelevant documents. We first defined five exclusion criteria:

\begin{itemize}[leftmargin=1.1em]
    \item EC1: The study is outside the scope of this study. It is not primarily related to source code comprehension, readability, or legibility, does not involve any comparison of different ways of writing code, neither direct nor indirect, or is clearly irrelevant to our research questions. For instance, we exclude studies focusing on high-level design, documentation, and dependencies (higher-level issues).
    \item EC2: The study is not a full paper (e.g., MSc dissertations, PhD theses, course completion monographs, short papers) or is not written in English: not considering these types of documents in systematic reviews is a common practice \cite{Kitchenham2015}. As a rule of thumb, we consider that full papers must be at least 5 pages long.
    \item EC3: The study is about readability metrics without an experimental evaluation.
    \item EC4: The study is about program comprehension aids, such as visualizations or other forms of analysis or sensory aids (e.g., graphs, trace-based execution, code summarization, specification mining, reverse engineering).
    \item EC5: The study focuses on accessibility, e.g., targets individuals with visual impairments or neurodiverse developers.
\end{itemize}

Each of the \nbPapersFromAllEngines documents was analyzed by an author of this paper, who checked the title and abstract of the document, and in some cases the methodology, against the exclusion criteria. The documents that do not meet any of the exclusion criteria were directly accepted to enter the next step (described in the next section). The documents that meet at least one exclusion criterion passed through a second round in the triage process, where each document was analyzed by a different author.
In the end of the two rounds, we discarded all documents that were annotated with at least one exclusion criterion in both rounds. We followed this two-round process to mitigate the threat of discarding potentially relevant studies in the triage. Because only rejected papers in the first round were analyzed in the second round, i.e., raters in the second round knew that the documents had been marked for exclusion, no inter-rater agreement analysis was conducted. We ended up with \nbPapersAfterExclusionCriteria documents.

\subsection{Initial Selection (Study Inclusion)}\label{sec:inclusion}

After discarding clearly irrelevant documents in the triage step, we applied the following inclusion criteria in the \nbPapersAfterExclusionCriteria papers to build our initial set of papers:

\begin{itemize}[leftmargin=1.1em]
    \item IC1 (Scope): The study must be primarily related to the topics of code comprehension, readability, legibility, or hard-to-understand code.
    \item IC2 (Methodology): The study must be or contain at least one empirical study, such as controlled experiment, quasi-experiment, case study, or survey \textit{involving human subjects}.
    \item IC3 (Comparison): The study must compare alternative programming constructs, coding idioms, or coding styles \textit{in terms of code readability or legibility}.
    \item IC4 (Granularity): The study must target fine-grained program elements and low-level/limited-scope programming activities. Not design or comments, but implementation.
\end{itemize}

The application of the inclusion criteria to a paper often requires reading not only the title and abstract as in the triage step, but also sections of introduction, methodology, and conclusion. If a given paper violates at least one inclusion criterion, the paper is annotated with ``not acceptable''. When there are doubts about the inclusion of a paper, the paper is annotated with ``maybe'' for further discussion. We also performed this step in two rounds, but differently from the triage, all papers in this step were independently analyzed by two different authors. We calculated the Kappa coefficient \cite{Cohen1960} for assessing the agreement between the two analyzes. We found $k = 0.323$, which is considered a fair agreement strength \cite{Kitchenham2015}.
In the end of this step, the papers annotated with ``acceptable'' in both rounds were directly selected, and papers annotated with ``not acceptable'' in both rounds were rejected. All the other cases were discussed by the four authors in live sessions to reach consensus.
We ended up with \nbPapersAfterInclusionCriteria papers.

\subsection{Study Quality Assessment}\label{sec:quality-assessment}

After the search, exclusion, and inclusion of papers, the remaining \nbPapersAfterInclusionCriteria papers passed through a final selection step, aiming to assess their quality. In this step, we aim to identify low-quality papers for removal.
To do so, we elaborated nine questions
that were answered for each paper. We adapted these questions from the work of Keele~\cite{keele2007guidelines}. 
There were three questions about study design, e.g., ``are the aims clearly stated?''; four questions about analysis, e.g., ``are the data collection methods adequately described?''; and two questions about rigor, e.g., ``do the researchers explain the threats to the study validity?''. There are three possible answers for each question: yes (1), partially (0.5), and no (0). The sum of the answers for a given paper is its score. The maximum is, therefore, 9. If a paper scores 0.5 or better in all questions, its overall score is 4.5 or more. Thus, we defined that a paper should score at least 4.5 to be kept in our list of papers.

Each paper was assessed by one of the authors. At the beginning of this step, each author selected one paper, performed the quality assessment, and justified to the other authors the given score for each question in a live discussion. This procedure allows us to align our understanding of the questions and avoid misleading assessments. The scores of the studies were: $min = 5$, $median = 8$, $max = 9$. Since the minimum score for keeping a paper is 4.5 and no paper scored less than 5, no studies were removed because of bad quality.

\vspace{5pt}
\noindent\textit{Inclusion of additional studies.}
Five relevant studies were not captured by our automatic search because our search string did not cover them, and they were unknown for us when we built the set of seed papers. We found some of these studies when we were reading already included studies in the final selection step---those studies were cited as ``related works'' to the ones included in our list.
The five studies were discussed by all the authors in a live session and, after reaching agreement about their pertinence, they were subjected to exclusion and inclusion criteria and quality evaluation.

\vspace{5pt}
\noindent\textit{Deprecated studies.}
We identified some papers that we refer to as \textit{deprecated}. A paper is deprecated if it was extended by another paper that we selected. For instance, the work of Buse and Weimer published in 2008~\cite{Buse2008} was extended in a subsequent paper~\cite{Buse2010}. In this case, we consider the former to be deprecated and only take the latter into account.

\subsection{Data Analysis}\label{sec:study_analysis}

In this step of the study, we analyzed a total of \nbPapersFinalSelection papers. Initially, we read all the papers in full and extracted from each one the data necessary to answer our research questions. This activity was carried out through a questionnaire that had to be filled in for each paper. For this extraction, the papers were divided equally among the researchers, and periodic meetings were held to discuss the extracted data. At this stage we seek to extract information about the characteristics of the studies, for example, whether they pertain to readability or legibility, the evaluation method, the tasks the subjects were required to perform, and information about the results.  

After extracting the data, we analyzed it so as to address the \numberToBeChecked{two} research questions. 
For the first research question, we collected all the tasks extracted from the \nbPapersFinalSelection studies. The first two authors examined these tasks together, identifying commonalities and grouping them together. 
All these tasks were then subject to discussion among all the authors. After we reached a consolidated list of tasks performed by the subjects of these studies, we organized them in three large groups, considering what is required from the subjects: (i) to provide information about the code; (ii) to act on the code; and (iii) to provide a personal opinion about the code. 

For the second research question, we adopted a similar procedure. Initially, we collected the extracted response variables. Since they were very diverse, we created groups that include multiple response variables with similar characteristics. For example, response variables related to correctness include the predicted output of a program, a recalled part of it, or a general description of its function. At the end, we elicited \numberToBeChecked{five} categories of response variables in the studies. The initial analysis of the data was conducted by the first two authors and later the results were refined with the collaboration of all the authors.

\vspace{5pt}
\noindent\textit{Data availability.}
Raw data, such as the list of the \nbPapersFromAllEngines documents returned by our automatic search, and additional detail about our methodology, such as the questionnaires we used for study quality assessment and data extraction, are available at \url{https://github.com/reydne/code-comprehension-review}.

\section{Results}\label{sec:results}

In this section we attempt to answer our \numberToBeChecked{two} research questions, based on the obtained data. 

\subsection{Tasks Performed by Human Subjects (RQ1)}\label{sec:result-tasks}

The essential code comprehension task is code reading. By construction, all the selected studies have at least one reading task where the subject is required to read a code snippet, a set of snippets, or even large, complete programs. Since all the studies compare two or more ways of writing code, the subjects often have multiple code reading tasks. In addition, the subjects are also expected to comprehend the code. However, there are different ways to measure subject comprehension performance. Subjects are asked to provide information about the code, act on the code, or provide personal opinion. In most studies, more than one kind of task was employed. \autoref{tab:tasks-types} summarizes the identified tasks.

A large portion of the primary studies, \numberToBeChecked{40} out of \nbPapersFinalSelection, required subjects to \textbf{provide information about the code}. For example, Benander et al.\cite{Benander1996} asked the subjects to explain using free-form text what a code snippet does, right after having read it. Blinman et al.\cite{Blinman2005} asked the subjects to choose the best description for a code snippet among multiple options. Explaining what the code does is not an objective method to evaluate comprehension because someone has to judge the answer. We found \numberToBeChecked{18} studies that employed this task.

The subjects of \numberToBeChecked{27} studies were asked to  answer questions about characteristics of the code. In some studies, the subjects were asked to predict the behavior of a source code just by looking at it. For example, Gopstein et al.\cite{Gopstein2017} and Ajami and Feitelson~\cite{Ajami2019} presented subjects with multiple code snippets and asked them to guess the outputs of these snippets.  
Dolado et al.\cite{Dolado2003} asked the subjects to answer a set of questions about expression results, final values of variables, and how many times loops were executed. In other studies, the subjects were asked questions about higher-level issues related to source code. For example, Scalabrino et al.\cite{Scalabrino2019} asked the subjects if they recognize an element of a specific domain or about the purpose of using an external component in the snippet (e.g., JDBC APIs).
Similarly, Binkley et al.\cite{Binkley2009} inquired the subjects about the kind of application or the industry where a specific line of code might be found. In addition, some studies required subjects to localize code elements of interest. 
For example, Binkley et al.\cite{Binkley2013} asked subjects to find identifiers in a snippet, marking each code line that has that identifier. Ceccato et al.\cite{Ceccato2009} asked the subjects to pinpoint the part of the code implementing a specific functionality.

Some studies made the assumption that code that is easy to understand is also easy to memorize. Therefore, they attempted to measure how much subjects remember the code. For example, Love~\cite{Love1977} asked subjects to memorize a program for three minutes and rewrite the program as accurately as possible in the next four minutes. Lawrie et al.~\cite{Lawrie2007} first presented a code snippet to the subjects. Then, in a second step, they listed six possible identifiers and the subjects had to select the ones that they recalled appearing in the code. Overall, \numberToBeChecked{seven} studies asked the subjects to remember the code.

On the other hand, some studies required the subjects to \textbf{act on the code}. In \numberToBeChecked{ten} studies subjects were asked to find and fix bugs in the code. Scanniello et al.~\cite{Scanniello2013} asked subjects to do so in two programs with different identifier styles. In \numberToBeChecked{eight} other studies the subjects were asked to modify the code of a working program, i.e., without the need to fix bugs.
For example, Jbara and Feitelson\cite{Jbara2014b} asked subjects to implement a new feature in a program seen in a previous task. Likewise, Schulze et al.\cite{Schulze2013} requested that subjects modify and delete annotated code with specific preprocessor directives, which also requires understanding the respective source code.

\input{table_tasks}

In a few studies, subjects were asked to write code from a description. Writing code \textit{per se} is not a comprehension task, but it may be associated to a comprehension task.
For example, Wiese et al.\cite{Wiese2019} first asked subjects to write a function that returns true if the input is 7 and false otherwise so that they could identify what code style (novice, expert, or mixed) the subjects preferred when coding. Afterwards, they asked the subjects to choose the most readable among three versions of a function, each one with a different code style. One of the goals of this study was to determine if the subjects write code in the same style that they find more readable.

Lastly, in \numberToBeChecked{30} studies subjects were asked to give their \textbf{personal opinion}. In \numberToBeChecked{nine} studies the subjects were inquired about their personal preferences or gut feeling without any 
additional task.
For example, Buse et al.\cite{Buse2010} asked them to rate (from 1 to 5) how legible or readable a code snippet is. Similarly, Arab\cite{Arab1992} asked subjects to classify the legibility of three presentation schemes of Pascal code in descending order. In the study of Santos and Gerosa~\cite{dosSantos2018}, the subjects chose the most readable between two functionally equivalent snippets. In other studies, \numberToBeChecked{21} in total, the subjects were asked about their personal opinion while they performed other tasks.
For example, O'Neal et al.\cite{ONeal1994} first asked subjects to read a code snippet, and then to state if they understood the snippet and to provide a description of its functionality. Similarly, Lawrie et al.\cite{Lawrie2007} asked subjects to provide a free-form written description of the purpose of a function and to rate their confidence in their description. In addition, subjects were asked to rate the difficulty of the comprehension tasks they had to perform in some studies, e.g., in the study of Fakhoury et al.\cite{Fakhoury2019b}.

\begin{summary}
To assess code readability and legibility, researchers conduct studies where subjects are asked to provide information about the code, act on the code, or give their personal opinion. We found out that \numberToBeChecked{16.7\%} of the studies only ask the subjects to provide personal opinion. Moreover, although most of the analyzed studies require subjects to provide information about source code, that information varies widely in nature. Subjects may be asked to predict the output of programs, identify code elements, or explain high-level functionality.
\end{summary}


\subsection{Response Variables (RQ2)}

\begin{figure}[t]
    \centering
    \includegraphics[trim=0 13 0 13,clip,width=0.45 \textwidth]{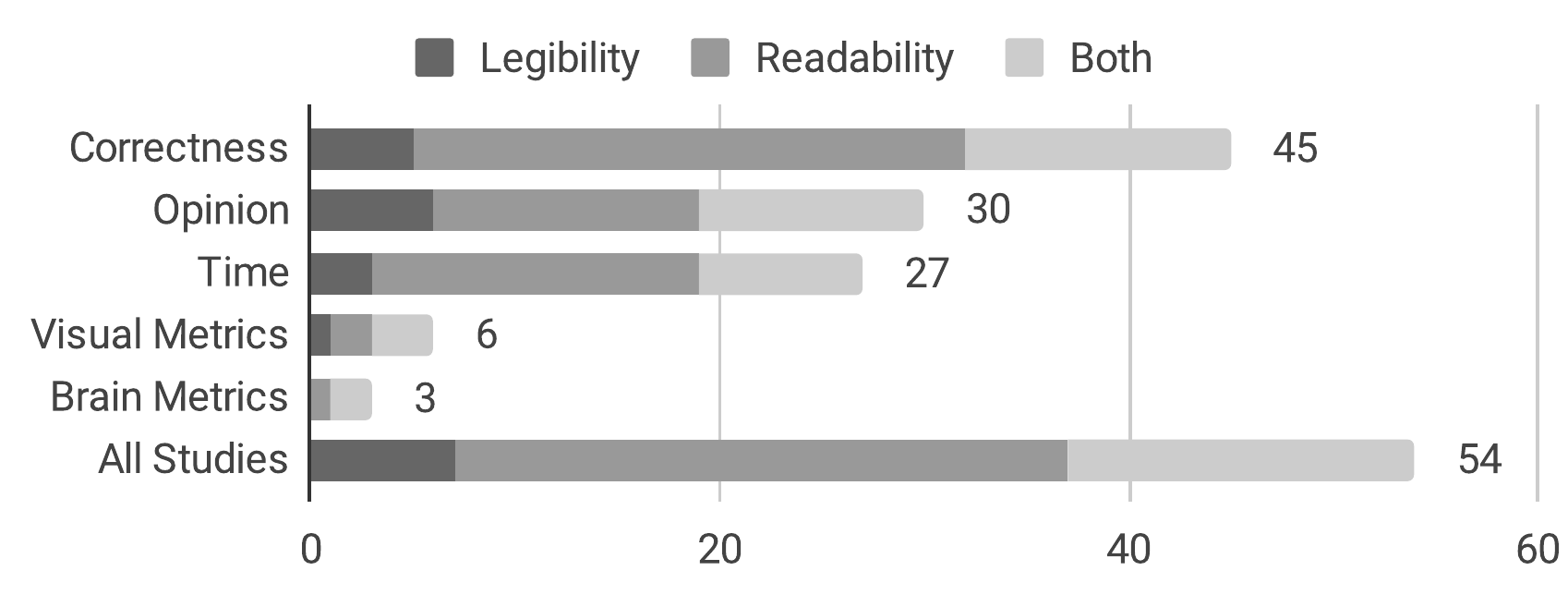}
    \caption{Frequency of response variables.}
    \label{fig:variables}
\end{figure}

Depending on the goals, methodology, and subjects of a study, response variables vary. This section presents the results obtained in the analysis of the response variables used in the selected studies. Since there is considerable diversity of response variables, we have organized them into \numberToBeChecked{five} categories.
\autoref{fig:variables} presents the frequency of all response variables. The numbers do not add up to \nbPapersFinalSelection, which is the overall number of analyzed studies, because most studies employed more than one response variable. In \autoref{tab:response}, we present a detailed synthesis of the identified response variables.

The performance of subjects was measured in some studies in terms of whether they were able to correctly provide information about programs just by looking at the source code. The response variables may pertain to code structure, semantics, use of algorithms, or program behavior, for instance. We aggregated response variables like these into a category called \textbf{correctness}. 
On the one hand, correctness can be objectively determined. For example, Bauer et al. \cite{Bauer2019} measured the subjects' code understanding by asking them to fill in a questionnaire with multiple-choice questions referring to program output. Gopstein et al.~\cite{Gopstein2017} and Ajami and Feitelson~\cite{Ajami2019} asked the subjects to predict the outputs of the execution of short programs. On the other hand, correctness was subjectively determined in some studies, where there was some margin to interpret if the results produced by a subject were correct.
For example, Blinman et al.~\cite{Blinman2005} evaluated if the subjects' textual description of a program was correct.
Similarly, Love~\cite{Love1977} asked subjects to write a textual description of a program and scored it based on a subjective evaluation. Overall, correctness response variables were employed by \numberToBeChecked{83.3\%} of the studies.

The \numberToBeChecked{second} most often employed response variable, present in \numberToBeChecked{30} studies, is the subjects' personal opinion. What is common to all these studies is the use of the preferences and gut feeling of the subjects, instead of the results of what they do, to assess readability and legibility. We grouped these response variables in a category called \textbf{opinion}. Scalabrino et al.~\cite{Scalabrino2019}, for example, asked the subjects to state whether they understood a code snippet or not. Santos and Gerosa~\cite{dosSantos2018} presented pairs of functionally equivalent snippets to the subjects and asked them to choose which one they think is more readable or legible. Stefik and Gellenbeck~\cite{Stefik2011} requested that subjects rate lists of words/symbols associated to programming constructs (e.g., conditionals and loops) based on how intuitive they think they are. Lawrie et al. \cite{Lawrie2007} asked the subjects to rate their confidence in their understanding of the code.

\input{table_response_variables}

The \textbf{time} that subjects spent to perform tasks was measured in multiple studies. This response variable category is the \numberToBeChecked{third} most often used in the analyzed studies, with \numberToBeChecked{27} instances. There is variety in the way the studies measured time.
For Ajami and Feitelson~\cite{Ajami2019}, \textit{``time is measured from displaying the code until the subject presses the button to indicate he is done''}.
Hofmeister et al.~\cite{Hofmeister2019} computed the time subjects spent looking at specific parts of a program. Geffen et al. \cite{Geffen2016} measured the response time for each question when subjects answer a multiple-choice questionnaire about the code. Instead of directly measuring time, 
Malaquias et al. \cite{Malaquias2017} counted the attempts to fix syntactic and semantic errors in the code.

In some studies, information about the process of the task was collected instead of its outcomes. In particular, multiple studies employed special equipment to track what the subjects see, and employed \textbf{visual metrics} as response variables. \numberToBeChecked{Six} of the analyzed studies employed some form of eye tracking.
For example, Blinkley et al. \cite{Binkley2013} computed the visual attention, measured as the amount of time during which a subject is looking at a particular area of the screen.
Bauer et al. \cite{Bauer2019} computed three visual metrics: fixation duration, fixation rate, i.e., the number of fixations per second, and saccadic amplitude, i.e., the spatial length of a saccade, that is, the transition between two fixations.
Hofmeister et al.~\cite{Hofmeister2019} employed a software tool that limits the subjects' view of the code to a few lines at a time. This frame can be shifted up and down using the arrow keys to reveal different parts of the code. This approach is called  ``letterboxing''. Hofmeister et al.\cite{Hofmeister2019} called each code frame an area of interest (AOI), and measured the time subjects spent on an AOI, first-pass reading times, and AOI visits.

Recently, some researchers have resorted to leveraging brain monitoring tools to understand what happens in the brain of the subjects during program comprehension. In total, \numberToBeChecked{three} of the analyzed studies employed response variables based on brain monitoring. Siegmund et al.~\cite{Siegmund2017} used functional magnetic resonance imaging (fMRI) to measure brain activity by detecting changes associated with blood flow. Fakhoury et al.~\cite{Fakhoury2019b} employed functional near-infrared spectroscopy (fNIRS) to measure brain activity through the hemodynamic response within physical structures of the brain. Yeh et al.~\cite{Yeh2017} leveraged electroencephalography (EEG) to monitor the electrical activity of the brain. These response variables were grouped into the \textbf{brain metrics} category.

The analyzed studies differ in the response variables they employed, depending on whether they aimed to assess readability, legibility, or both. As shown in \autoref{fig:variables}, readability studies leveraged all the response variable categories whereas no legibility study leveraged brain metrics. This is not surprising, considering the much lower number of legibility studies.
In addition, a high proportion of legibility studies (\numberToBeChecked{86\%}) employed opinion response variables. This is not the case for studies only about readability (\numberToBeChecked{43\%}) or about both (\numberToBeChecked{65\%}).

Most studies, \numberToBeChecked{38} in total, employed more than one variable to validate their hypotheses. The response variables time and correctness are the ones that most often appear together (\numberToBeChecked{27} studies), followed by opinion and correctness (\numberToBeChecked{21} studies).
In a different manner, \numberToBeChecked{16} studies employed only one response variable, and correctness and opinion were the response variables employed in isolation. Among these, \numberToBeChecked{9} studies (\numberToBeChecked{16.7\%}) employed only personal opinion. The use of this response variable category in isolation is a clear threat to the validity of these studies. Furthermore, no study used time as the only response variable. This makes sense since, even though time is often used as a proxy for effort, it is not significant by itself.

\begin{summary}
There are \numberToBeChecked{five} categories of response variables.  Correctness is the most widely used, being employed in \numberToBeChecked{83.3\%} of the analyzed studies. Time and opinion are also often employed (\numberToBeChecked{50\%} and \numberToBeChecked{55.6\%} of the studies, respectively).
A significant number of studies used the variable time and/or opinion associated with correctness. On the other hand, \numberToBeChecked{30\%} of the studies employed a single response variable.
The readability and readability+legibility studies used all the response variable categories while almost all the legibility studies used the opinion response variable.
\end{summary}

\section{Program Comprehension as a Learning Activity}\label{sec:program-comprehension-as-a-learning-activity}

The empirical studies analyzed in this work involve a wide range of tasks to be performed by their subjects (see \autoref{sec:result-tasks}). For example, they may ask subjects to memorize a program, follow its execution step by step, answer questions about it, or write a high-level explanation of what it does.  All these tasks are conducted with the goal of evaluating readability and legibility. However, they demand different cognitive skills from the subjects and, as a consequence, evaluate different aspects of readability and legibility. We attempt to shed a light on this topic by analyzing the cognitive skill requirements associated with each kind of task. 

According to the Merriam-Webster Thesaurus, to learn something is \textit{``go gain an understanding of''} it. We follow this definition by treating the problem of program comprehension (or understanding) as a learning problem. 
In this section we propose an adaptation of the learning taxonomy devised by Fuller and colleagues~\cite{fuller2007developing} to the context of program comprehension. This taxonomy is itself an adaptation of Bloom's taxonomy of educational objectives~\cite{bloom1956taxonomy} to the context of Software Development. Fuller et al.~\cite{fuller2007developing} state that \textit{``learning taxonomies [..] describe the learning stages at which a learner is operating for a certain topic.''} A learning taxonomy supports educators in establishing intended learning outcomes for courses and evaluate the students' success in meeting these goals. According to Biggs~\cite{Biggs:1999:TQL} and Fuller et al.~\cite{fuller2007developing}, learning taxonomies can help with ``understanding about understanding'' and 
``communicating about understanding''. Based on this description, they seem to be a good fit to help researchers better understand studies about program understanding and communicate about them. Our central idea is to use the elements defined by the taxonomy of Fuller et al.~\cite{fuller2007developing}, with some adaptions, to identify and name the cognitive skills required by different tasks employed by code readability and legibility studies.

\begin{table*}[t]
{\scriptsize
    \caption{Learning activities extended from Fuller et al. \cite{fuller2007developing}---Inspect and Memorize are not in the original taxonomy. Opinion is not included because it is not directly related to learning.}
    \label{tab:study-analysis-activities}
    \centering
    \begin{tabular}{@{} p{0.05\textwidth} p{0.44\textwidth} p{0.457\textwidth} @{}}
        \toprule
        Activity & Description & Example \\
        \midrule
        Adapt & Modify a solution for other domains/ranges. This activity is about the modification of a solution to fit in a given context. & Remove preprocessor directives to reduce variability in a family of systems~\cite{Schulze2013}. \\
        
        Analyze & Probe the [time] complexity of a solution. & Identify the function where the program spends more time running.\\
       
        Apply & Use a solution as a component in a larger problem. Apply is about changing a context so that an existing solution fits in it. & Reuse of off-the-shelf components. \\
          
        Debug & Both detect and correct flaws in a design. & Given a program, identify faults and fix them~\cite{Scanniello2013}. \\
        
        Design & Devise a solution structure. The input to this activity is a problem specification. & Given a problem specification, devise a solution satisfying that specification. \\
        
        Implement & Put into lowest level, as in coding a solution, given a completed design. & Write code using the given examples according to a specification~\cite{Stefik2013}. \\
        
        Model & Illustrate or create an abstraction of a solution. The input is a design. & Given a solution, construct a UML model representing it. \\
       
        Present & Explain a solution to others. & Read a program and then write a description of what it does and how~\cite{Chaudhary1980}. \\
        
        Recognize & Base knowledge, vocabulary of the domain. In this activity, the subject must identify a concept or code structure obtained before the task to be performed. & ``The sorting algorithm (lines 46-62) can best be described as: (A) bubble sort (B) selection sort (C) heap sort (D) string sort (E) partition exchange. sort''~\cite{Tenny1988}. \\
       
        Refactor & Redesign a solution (as for optimization).
        The goal is to modify non-functional properties of a program or, at a larger scale, reengineer it. & Rewrite a function so as to avoid using conditional expressions.  \\
       
        Relate & Understand a solution in context of others. This activity is about identifying distinctions and similarities, pros and cons of different solutions. & Choose one out of three high-level descriptions that best describe the function of a previously studied application~\cite{Blinman2005}. \\
        
        Trace & Desk-check a solution. Simulate program execution while looking at its code. & Consider the fragment ``x=++y'': what is the final value of x if y is -10?~\cite{Dolado2003}. \\
        
        Inspect* & Examine code to find or understand fine-grain static elements. Inspect is similar to Analyze, but it happens at compile time instead of run time. & ``All variables in this program are global.'' [true/false] \cite{Miara1983}. \\

        Memorize* & Memorize the code in order to reconstruct it later, partially or as a whole. & Given a program, memorize it in 3 minutes and then reconstruct it in 4~\cite{Love1977}. \\
        
        \bottomrule
    \end{tabular}
    \vspace{-8pt}
    }
\end{table*}

\subsection{A Learning Taxonomy}

Bloom's taxonomy \cite{bloom1956taxonomy} consists of three hierarchical models aiming to classify educational learning objectives in terms of complexity and specificity. In the context of this work, the cognitive domain of this taxonomy is the most relevant. It recognizes that learning is a multi-faceted process that requires multiple skills with different levels of complexity and which build upon each other. This taxonomy has been later revised by Anderson et al.~\cite{bloomrevised2001}. The most visible aspect of the revised taxonomy is a list of six activities that define progressively more sophisticated levels of learning: remember, understand, apply, analyze, evaluate, and create. 

Fuller et al.~\cite{fuller2007developing} proposed an adaptation of the revised version of Bloom's taxonomy for the area of Computer Science, with a particular emphasis on software development. They developed a set of activities that build upon Bloom's revised taxonomy and organized them in a model that emphasizes that some activities in software development involve acting on knowledge, instead of just learning. We leverage this taxonomy and apply it in the context of program comprehension. \autoref{tab:study-analysis-activities} presents the activities (i.e., cognitive skills) devised by Fuller et al.~\cite{fuller2007developing}. Also, we introduce two activities (marked with ``*'') that stem directly from tasks performed by subjects in some of our primary studies and require skills that are not covered by the original set of activities. \autoref{tab:study-analysis-activities} also presents examples of the activities, which are either extracted from tasks from our primary studies or general examples when no task involved that activity.
In the next section we leverage these activities to gain a better understanding of the tasks conducted by subjects in our primary studies.

\subsection{Mapping Tasks to Learning Activities}\label{sec:mapping}

We analyzed the tasks that subjects performed in the primary studies (\autoref{sec:result-tasks}) and identified which activities (\autoref{tab:study-analysis-activities}) they required.
A task can require subjects to conduct one or more activities. For example, Avidan and Feitelson\cite{Avidan2017} asked subjects to explain what a function does. This is an instance of the Present activity. Miara et al.\cite{Miara1983} applied a questionnaire that asked subjects to identify specific code elements and also inquired them about the results of expression evaluations. In this case, both Inspect and Trace activities were executed. Besides the activities in \autoref{tab:study-analysis-activities}, we also considered Giving an Opinion (hereafter, ``Opinion''), which is not part of the taxonomy because it is not a learning activity. We considered Opinion to reflect the intrinsically biased and unreliable nature of tasks that ask subjects to provide their opinion. Previous work \cite{Rossbach:2010:ITP,Castor:2011:STM} has shown that evidence in software engineering studies often contradicts opinions. This kind of analysis falls outside the scope of this paper.

\begin{figure}[t] 
    \centering
    \includegraphics[width=0.40 \textwidth]{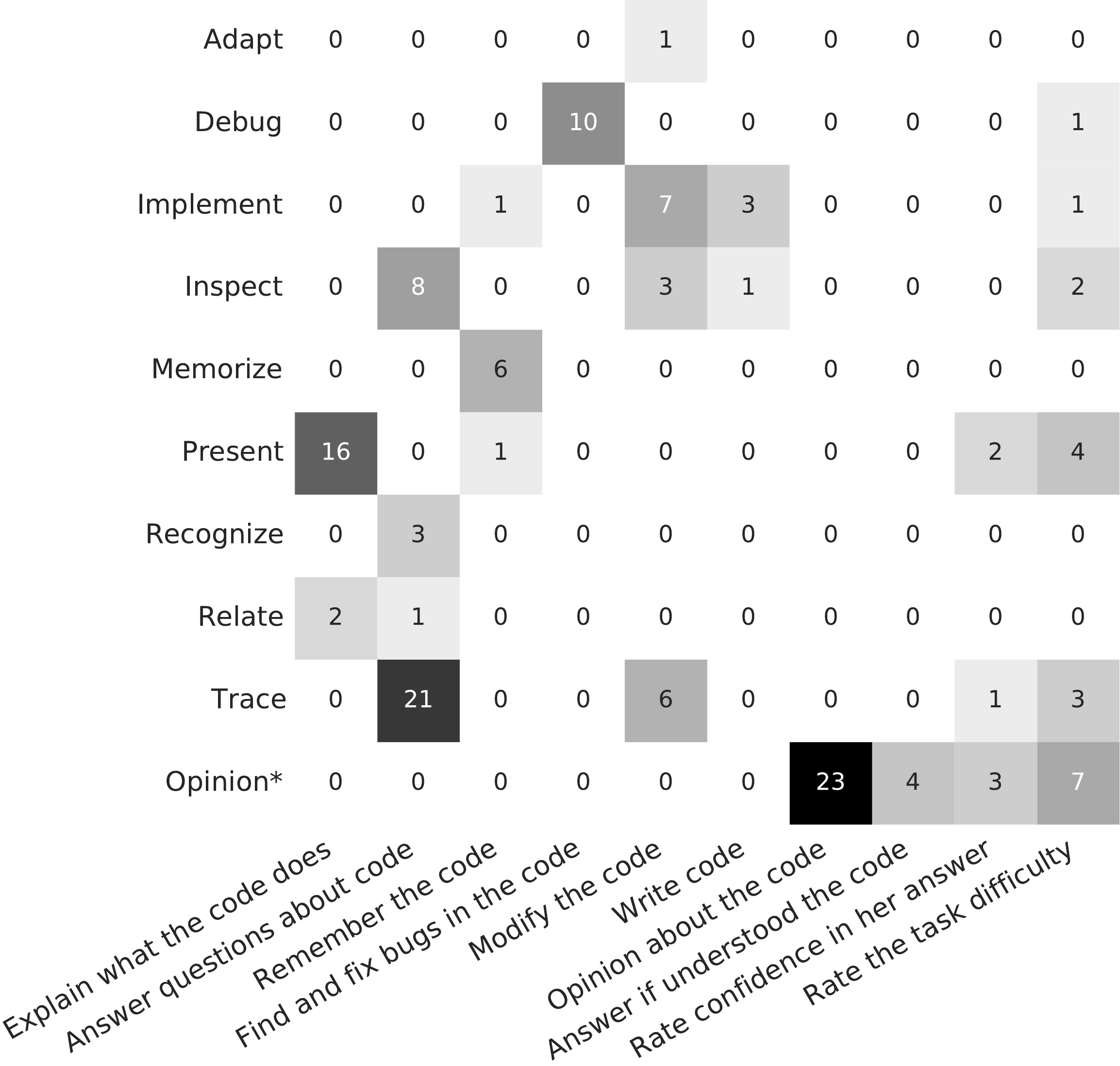}
    \caption{Confusion matrix of tasks (columns) and learning activities (rows).}
    \label{fig:tasks_activities_matrix}
\end{figure}

\autoref{fig:tasks_activities_matrix} presents a confusion matrix to show the frequency of studies in which tasks and learning activities co-occur. For instance, there are six studies that involves the ``Remember the code'' task in the context of the Memorize learning activity. The matrix shows that there is a direct correspondence between some tasks and activities. For example, all the instances of the ``Find and fix bugs in the code'' task involve the Debug activity, and all the tasks that require subjects to provide an opinion are connected to the Opinion activity. In addition, some tasks may be connected to various activities. For instance, ``Modify the code'' may require subjects to Implement, Trace, Inspect, or Adapt the code to be modified. This makes sense; to modify a program, one may have to understand its static elements and its behavior, as well as adapt code elements to be reused. Another example is ``Answer questions about code'', which often requires subjects to Trace and Inspect code. Furthermore, ``Explain what the code does'' is usually related to Present. Notwithstanding, in two studies~\cite{ONeal1994,Blinman2005} subjects were presented with multiple descriptions for the same code snippet and asked which one is the most appropriate. This requires the subject to Relate different programs and descriptions. Finally, there are some nonintuitive relationships between tasks and activities. Chaudhary~\cite{Chaudhary1980} asked the subjects to reconstruct a program after spending some time looking at it. This is a ``Remember the code'' task where the subjects have to explain what the program does by writing a similar program. It is a mix of Present and Implement. In another example, Lawrie et al.~\cite{Lawrie2007} asked the subjects to describe what a code snippet does and rate their confidence in their answers. Rating one's confidence in an answer is about Opinion but it also must be associated with some other activity, since some answer must have been given in the first place. In this case, the activity is to Present the code snippet, besides giving an Opinion. A similar phenomenon can be observed for studies where subjects were asked to ``Rate the task difficulty'' they had performed, e.g., \cite{Jbara2017,Fakhoury2019b}.

\begin{figure}[t] 
    \centering
    \includegraphics[trim=0 20 0 20,clip,width=0.45 \textwidth]{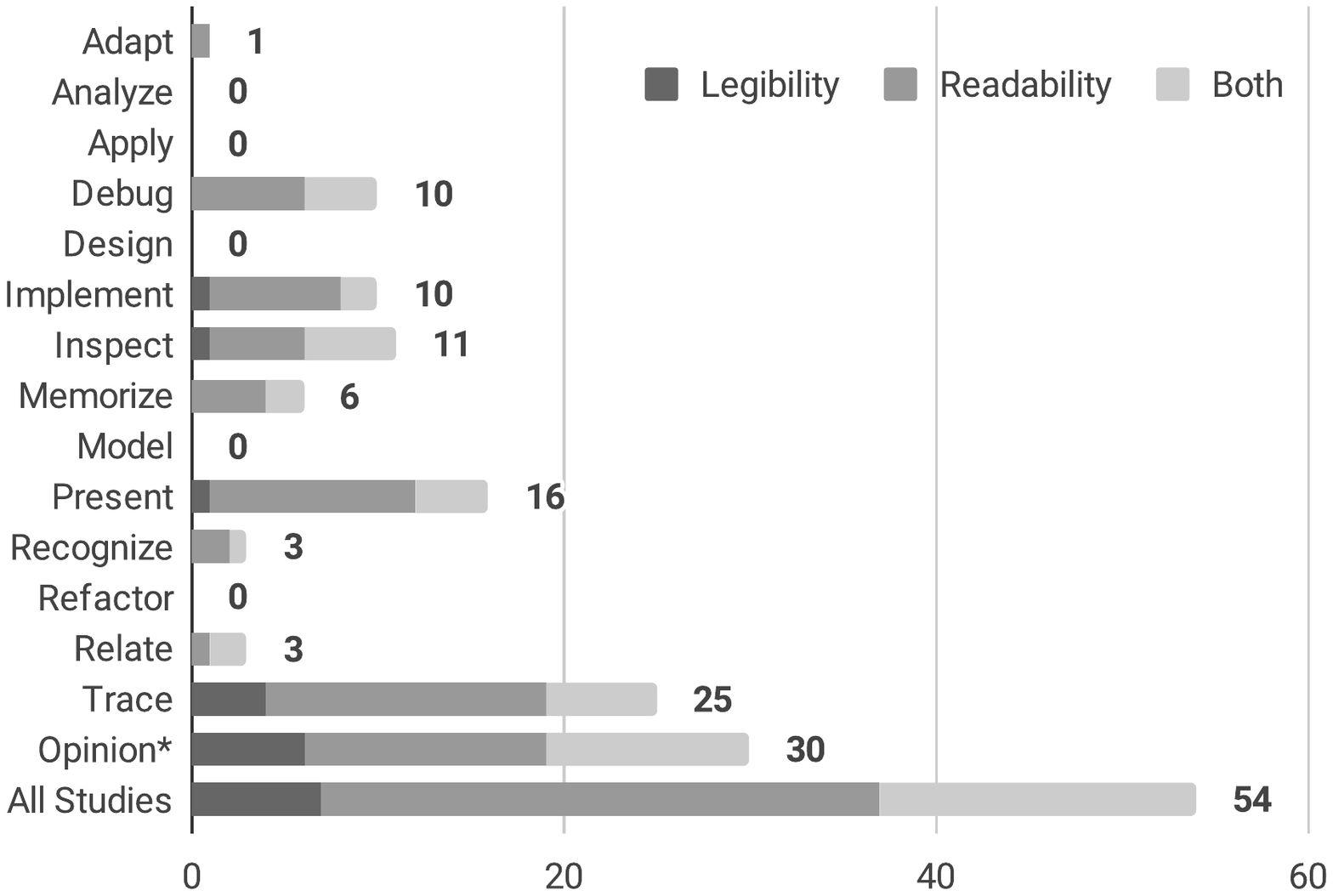}
    \caption{Frequency of learning activities and subject opinion.}
    \label{fig:activities_frequency}
    \vspace{-12pt}
\end{figure}

\autoref{fig:activities_frequency} shows the frequency of the activities required in the analyzed studies, separating readability and legibility studies, as well as those targeting both attributes. Opinion was required by more studies than any learning activity.
The two most widely used learning activities require that subjects extract information from programs. The most widely used one, Trace, was required in \numberToBeChecked{25} studies, where subjects were asked to predict the output value of a program or the value of a variable. Coming in second, the Present activity was required in \numberToBeChecked{16} studies, where usually subjects were asked to explain what the program does. It is also common for subjects to be asked to Inspect the code (\numberToBeChecked{11} studies), so as to determine, e.g., how many loops are there in the program. Subjects were required to Implement or Debug code in \numberToBeChecked{10} studies each. 
\autoref{fig:activities_frequency} also highlights that studies focusing solely on legibility and on readability and legibility combined require subjects to give Opinion more than any learning activity. Even considering the small number of legibility studies, this may be a hint that researchers have yet to figure out the effective ways to meaningfully assess legibility-related attributes.

\begin{figure}[t] 
    \centering
    \includegraphics[width=0.37\textwidth]{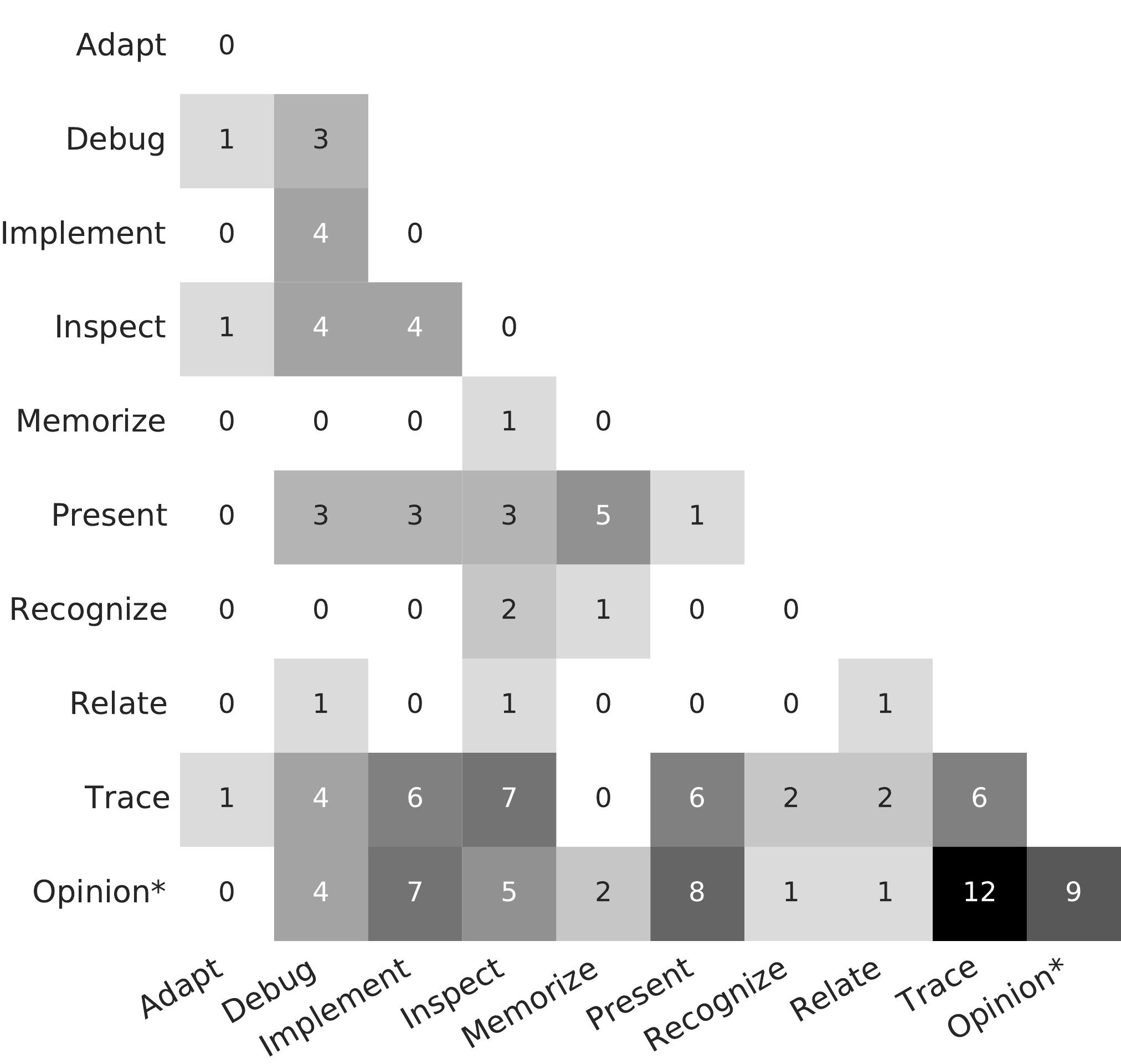}
    \caption{Co-occurrence of learning activities and subject opinion---the main diagonal represents the number of studies that an activity or opinion is used alone.}
    \label{fig:activities_matrix}
\end{figure}

\autoref{fig:activities_matrix} highlights how many studies use combinations of two activities.
More than one activity was employed by \numberToBeChecked{34} of the studies, which combine mainly the most popular activities, i.e., Trace, Present, and Inspect. The numbers in the diagonal indicate how many studies required their subjects to use a single activity. In total, \numberToBeChecked{20} studies employed a single activity. This number amounts to \numberToBeChecked{37\%} of our primary studies. The use of a single learning activity in an evaluation of readability or legibility suggests a narrow focus, since these attributes are multi-faceted, as highlighted by \autoref{tab:study-analysis-activities}. 

\subsection{A Two-Dimensional Model for Program Comprehension}\label{sec:2d}

\begin{figure}[t] 
    \centering
    \includegraphics[clip,width=0.488 \textwidth]{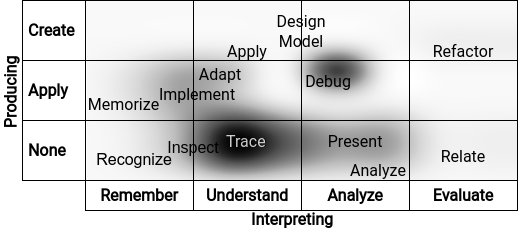}
    \caption{Learning activities and their frequency.}
    \label{fig:heatmap_activities}
\end{figure}

Besides the list of activities, Fuller et al.~\cite{fuller2007developing} proposed a taxonomy adapted from the revised version of Bloom's taxonomy \cite{bloomrevised2001}. It is represented by two semi-independent dimensions, Producing and Interpreting. Each dimension defines hierarchical linear levels where a deeper level requires the competencies from the previous ones. Producing has three levels (None, Apply, and Create) and Interpreting has four (Remember, Understand, Analyze, and Evaluate).  \autoref{fig:heatmap_activities} represents this two-dimensional model.
According to Fuller et al.~\cite{fuller2007developing}, a level appearing more to the right of the figure along the Interpreting dimension (x-axis), e.g., Evaluate, requires more competencies than one appearing more to the left, e.g., Remember. The same applies to the levels appearing nearer the top along the Producing dimension (y-axis).

The activities in \autoref{tab:study-analysis-activities} were positioned in \autoref{fig:heatmap_activities} in conformance to their required competencies (we did not consider Opinion as it is not part of the learning taxonomy).
We complemented the two-dimensional model by including the two activities that we introduced, Inspect and Memorize. \autoref{fig:heatmap_activities} is a heatmap that presents how frequently each competence was required by the studies analyzed in this work. Dark regions are associated to more frequent activities.

The Interpreting dimension indicates that most activities employed in studies evaluating readability and legibility occur at the Understand level, followed by Analyze and, to a lesser extent, Remember. The higher competence level of the Interpreting dimension, Evaluate, is almost never used. Considering the Producing dimension, where the ability to design and build a new product, e.g., a program, is evaluated, we notice that None is the most representative, followed by Apply. This is to be expected since most of the studies
focus on comprehension-related activities.
However, program comprehension is often required for work that involves the Apply and Create levels. 
Nevertheless, these levels are rarely tackled by our primary studies. This reinforces the point raised in \autoref{sec:mapping}, that many of these studies have a narrow evaluation focus. 

\begin{figure}[t]
\begin{subfigure}{.158\textwidth}
  \centering
  \includegraphics[width=1\linewidth]{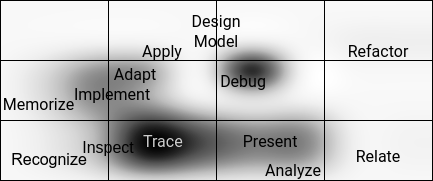}  
  \caption{Readability}
  \label{fig:heatmap_readability}
\end{subfigure}
\begin{subfigure}{.158\textwidth}
  \centering
  \includegraphics[width=1\linewidth]{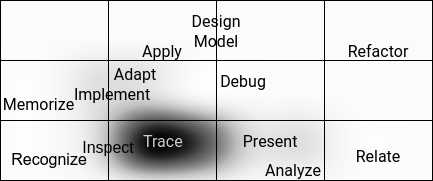}  
  \caption{Legibility}
  \label{fig:heatmap_legibility}
\end{subfigure}
\centering
\begin{subfigure}{.158\textwidth}
  \centering
  \includegraphics[width=1\linewidth]{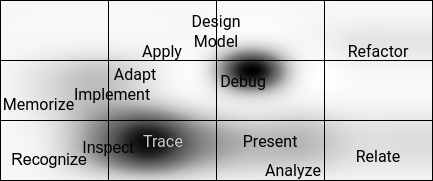}  
  \caption{Both}
  \label{fig:heatmap_both}
\end{subfigure}
\caption{Learning activities by readability and legibility. This figure decomposes \autoref{fig:heatmap_activities} to show spatial disposition.}
\label{fig:legibility_and_readability_heatmap}
\end{figure}

We built three additional heatmaps to emphasize the differences between readability and legibility studies when considering the two-dimensional model. \autoref{fig:legibility_and_readability_heatmap} presents the heatmaps for readability studies, legibility studies, and both studies in sequence. \autoref{fig:heatmap_readability} is very similar to \autoref{fig:heatmap_activities}. 
In contrast, \autoref{fig:heatmap_legibility} shows that legibility studies are concentrated in the Understand level, where Trace is the main activity. Finally, \autoref{fig:heatmap_both} presents the heatmap for studies that tackle both readability and legibility together. Albeit similar to \autoref{fig:heatmap_activities}, the area in the intersection between the Analyze level of the Interpreting dimension and the Apply level of the Producing dimension is darker in this figure. This reflects the proportionally higher number of studies that employ the Debug activity. 

\section{Threats to Validity}\label{sec:threats-to-validity}

\textit{Construct validity}. Our study was built on the selected primary studies, which stem from the search and selection processes. The search for studies relies on a search string, which was defined based on the seed papers. We chose only conferences to search for seed papers. A paper published in a journal is often an extension of a conference paper and, in Computer Science, the latest research is published in conferences. Furthermore, we focused on conferences only as a means to build our search string. Journal papers were considered in the actual review.
Additionally, we only used three search engines for our automatic search. Other engines, such as Springer and Google Scholar, could return different results. However, the majority of our seed studies were indexed by ACM and IEEE, and then we used Scopus to expand our search. Moreover, while searching on the ACM digital library, we used the \textit{Guide to the Computing Literature}, which retrieves resources from other publishers, such as Springer. Finally, we avoided Google Scholar because it returned more than 17 thousand documents for our search string. 

\textit{Internal validity}. This study was conducted by four researchers. We understand that this could pose a threat to its internal validity since each researcher has a certain knowledge and way of conducting her research activities. However, each researcher conducted her activities according to the established protocol, and periodic discussions were conducted between all researchers. Another threat to validity is the value of Cohen's Kappa in the study inclusion step ($k = 0.323$), which is considered fair. This value stems from the use of three possible evaluations (``acceptable'', ``not acceptable'', and ``maybe'') in that step. However, we employed ``maybe'' to avoid having to choose between ``acceptable'' and ``not acceptable'' when we had doubts about the inclusion of a paper---all papers marked with at least one ``maybe'' were discussed between all authors.
Moreover, a few primary studies do not report in detail the tasks the subjects performed. Because of that, we might have misclassified these studies when mapping studies to task types.

\textit{External validity}. Our study focuses on studies that report on comparisons of alternative ways of writing code, considering low-level aspects of the code. Our findings might not apply to other works that evaluate code readability or legibility.

\textit{Conclusion validity}. According to Kitchenham et al. \cite{Kitchenham2015}, conclusion validity is concerned with how reliably we can draw conclusions about the relationship between a treatment and the outcomes of an empirical study. In a systematic review, that relates to the data synthesis and how well this supports the conclusions of the review. The threats of our study related to it are therefore presented as internal validity.

\section{Conclusion}\label{sec:conclusion}

We presented a systematic literature review on how code readability and legibility are evaluated in human-centric studies that compare different ways of writing equivalent code. Our goal was to investigate what tasks are performed by subjects and what response variables are employed in those studies, as well as what cognitive skills are required by those tasks. We presented comprehensive classifications for both tasks and response variables.
Moreover, we adapted a learning taxonomy to program comprehension, mapping the tasks identified in the literature review to the cognitive skills that comprise the taxonomy. This study highlighted limitations of primary studies: (i) \numberToBeChecked{37\%} of them exercised a single cognitive skill; (ii) \numberToBeChecked{16.7\%} only employed personal opinion as a response variable; and (iii) few studies evaluated readability and legibility in a way that simulates real-world scenarios, where program comprehension is part of a more complex task, requiring higher-level cognitive skills.
We also introduced a separation between code readability and legibility. This separation is common in other areas such as linguistics, but, to the best of our knowledge, we are the first to propose it in the context of software engineering.
We are performing a systematic literature review focusing on the different ways of writing code and which ones improve or hinder readability and legibility.

\noindent\textbf{Acknowledgements.} We thank the anonymous reviewers for their valuable feedback on this paper. This research was partially funded by CNPq/Brazil (304755/2014-1, 406308/2016-0, 465614/2014-0), and FACEPE/Brazil (APQ-0839-1.03/14, IBPG-0690-1.03/19, 0388-1.03/14, 0592-1.03/15).

\clearpage

\balance

\bibliographystyle{IEEEtran}
\bibliography{references}

\end{document}

%% file: table_tasks.tex
\begin{table}[t]
{\scriptsize
\caption{Task types and their corresponding studies. A study may involve more than one type of task.}
\label{tab:tasks-types}
\centering
\begin{tabular}{@{} p{0.175\textwidth} p{0.29\textwidth} @{}}
\toprule
Task type            & Studies                                                                                                                                                                                                                                                                                                                                                                                                                    \\ \midrule
\multicolumn{2}{@{}l}{\textbf{\underline{provide information about the code (\numberToBeChecked{40} studies)}}}                                                                                                                                                                                                                                                                                                                                                                                                                                    \\
explain what the code does               & \numberToBeChecked{18} studies: \cite{Avidan2017,Benander1996,Beniamini2017,Binkley2013,Boysen1980,Chaudhary1980,Fakhoury2019b,Jbara2014b,Jbara2017,Kasto2013,Lawrie2007,Love1977,Miara1983,Wiedenbeck1986,Wiedenbeck1991,Woodfield1981,Blinman2005,ONeal1994}                                                                                                                   \\
answer questions about code characteristics         & \numberToBeChecked{27} studies: \cite{Ajami2019,Bauer2019,Boysen1980,Dolado2003,Geffen2016,Gopstein2017,Iselin1988,Kasto2013,Mackowiak2018,ONeal1994,Schulze2013,Siegmund2017,Sykes1983,Teasley1994,Trockman2018,Wiese2019,Wiese2019b,Woodfield1981,Yeh2017,Binkley2009,Binkley2013,Scalabrino2019,Miara1983,Tenny1988,Ceccato2009,Oman1988,Oman1990} 
\\
remember (part of) the code & \numberToBeChecked{7} studies: \cite{Binkley2009,Binkley2013,Lawrie2007,Love1977,Wiedenbeck1986,Wiedenbeck1991,Chaudhary1980}                                                                                                                                                                                                                                                                                                           \\ \midrule
\multicolumn{2}{@{}l}{\textbf{\underline{act on the code (\numberToBeChecked{15} studies)}}}                                                                                                                                                                                                                                                                                                                                                                                                                                                       \\
find and fix bugs in the code  & \numberToBeChecked{10} studies: \cite{Beniamini2017,Jbara2014b,Kleinschmager2012,Malaquias2017,Scanniello2013,Schankin2018,Fakhoury2019b,Hofmeister2019,Schulze2013,Siegmund2017}                                                                                                                                                                                                                                   \\
modify the code                & \numberToBeChecked{8} studies: \cite{Beniamini2017,Kleinschmager2012,Ceccato2009,Geffen2016,Malaquias2017,Jbara2014b,Schulze2013,Wiese2019b}                                                                                                                                                                                                                                              \\
write code                 & \numberToBeChecked{3} studies: \cite{Stefik2013,Wiese2019,Wiese2019b}                                                                                                                                                                                                                                                                                                                                                                                               \\ \midrule
\multicolumn{2}{@{}l}{\textbf{\underline{provide personal opinion (\numberToBeChecked{30} studies)}}}                                                                                                                                                                                                                                                                                                                                                                                                                                              \\
opinion about the code (readability or legibility)                         & \numberToBeChecked{23} studies: \cite{Arnaoudova2016,Buse2010,Jbara2014b,Jbara2017,Oman1988,Oman1990,ONeal1994,Posnett2011,Scalabrino2018,Stefik2011,Stefik2013,Arab1992,Bauer2019,Avidan2017,Beniamini2017,dosSantos2018,Medeiros2019,Wiese2019,Wiese2019b,Ceccato2009,Malaquias2017,Sykes1983,Wang2014}                                                                         \\
answer if understood the code       & \numberToBeChecked{4} studies: \cite{ONeal1994,Scalabrino2019,Sykes1983,Trockman2018}                                                                                                                                                                                                                                                                                                                                                                                         \\
rate confidence in her answer & \numberToBeChecked{3} studies: \cite{Lawrie2007,Wiedenbeck1991,Yeh2017}                                                                                                                                                                                                                                                                                                                                                                                                            \\
rate the task difficulty                    & \numberToBeChecked{7} studies: \cite{Fakhoury2019b, Miara1983, Bauer2019, Ceccato2009, Jbara2014b, Jbara2017, Yeh2017}                                                                                                                                                                                                                                                                                                                                                                                               \\ \bottomrule
\end{tabular}
}
\end{table}

%% file: table_response_variables.tex
\begin{table}[t]
{\scriptsize
\caption{Response variables and their corresponding studies.}
    \label{tab:response}
    \centering
    \begin{tabular}{@{} p{0.12\textwidth} p{0.06\textwidth} p{0.26\textwidth} @{}}
        \toprule
        Category and type & Sub-type & Studies \\
        \midrule
        \multicolumn{3}{@{}l}{\textbf{\underline{Correctness (\numberToBeChecked{45} studies)}}}        \\
        Objective & Binary & \numberToBeChecked{37} studies:
        \cite{Ajami2019,Bauer2019,Binkley2009,Binkley2013,Blinman2005,Boysen1980,Ceccato2009,Dolado2003,Fakhoury2019b,Geffen2016,Gopstein2017,Hofmeister2019,Kleinschmager2012,Iselin1988,Jbara2014b,Lawrie2007,Love1977,Mackowiak2018,Malaquias2017,Miara1983,Oman1988,Oman1990,ONeal1994,Scalabrino2019,Scanniello2013,Schankin2018,Siegmund2017,Sykes1983,Teasley1994,Tenny1988,Trockman2018,Wiedenbeck1986,Woodfield1981,Wiedenbeck1991,Wiese2019,Wiese2019b,Yeh2017} \\
        {} & Scale & \numberToBeChecked{3} studies:
        \cite{Schulze2013,Stefik2013,Beniamini2017} \\
        Subjective & Binary & \numberToBeChecked{11} studies:
        \cite{Avidan2017,Benander1996,Boysen1980,Chaudhary1980,Fakhoury2019b,Kasto2013,Miara1983,Wiedenbeck1986,Wiedenbeck1991,Wiese2019,Wiese2019b} \\
        {} & Rate & \numberToBeChecked{7} studies:
        \cite{Beniamini2017,Jbara2017,Lawrie2007,Binkley2013,Jbara2014b,Love1977,Woodfield1981} \\
        
        \midrule
        \multicolumn{3}{@{}l}{\textbf{\underline{Opinion (\numberToBeChecked{30} studies)}}}        \\
        Personal preference & Rate & \numberToBeChecked{9} studies: \cite{Arnaoudova2016,Buse2010,Medeiros2019,Oman1988,Oman1990,Posnett2011,Scalabrino2018,Stefik2011,Stefik2013} \\
        {} & Choice & \numberToBeChecked{6} studies: \cite{Avidan2017,Beniamini2017,dosSantos2018,Wang2014,Wiese2019,Wiese2019b} \\
        {} & Ranking & \numberToBeChecked{1} study: \cite{Arab1992} \\
        On understandability & Binary & \numberToBeChecked{2} studies: \cite{Scalabrino2019,Trockman2018} \\
        {} & Rate & \numberToBeChecked{2} studies: \cite{Sykes1983,ONeal1994} \\
        Professional opinion & Acceptability & \numberToBeChecked{2} studies: \cite{Medeiros2019,Malaquias2017} \\
        \multirow{1}{*}{Rate answer confidence} & Rate & \numberToBeChecked{3} studies: \cite{Lawrie2007,Wiedenbeck1991,Yeh2017} \\
        Rate task difficulty & Ranking & \numberToBeChecked{1} study: \cite{Bauer2019} \\
        {} & Rate & \numberToBeChecked{6} studies: \cite{Ceccato2009,Fakhoury2019b,Jbara2017,Jbara2014b,Miara1983,Yeh2017} \\
        
        \midrule
        \multicolumn{3}{@{}l}{\textbf{\underline{Time (\numberToBeChecked{27} studies)}}}        \\        
        \multicolumn{2}{@{}l}{Time to complete task} & \numberToBeChecked{20} studies: \cite{Ajami2019,Bauer2019,Benander1996,Beniamini2017,Blinman2005,Ceccato2009,Dolado2003,Hofmeister2019,Iselin1988,Jbara2014b,Jbara2017,Kleinschmager2012,Malaquias2017,ONeal1994,Scanniello2013,Schankin2018,Oman1990,Oman1988,Schulze2013,Binkley2013} \\
        \multicolumn{2}{@{}l}{Time reading code} & \numberToBeChecked{5} studies: \cite{Avidan2017,Binkley2013,Boysen1980,Fakhoury2019b,Scalabrino2019} \\
        \multicolumn{2}{@{}l}{Time per question} & \numberToBeChecked{4} studies: \cite{Binkley2009,Geffen2016,Boysen1980,Mackowiak2018} \\
        \multicolumn{2}{@{}l}{Number of attempts} & \numberToBeChecked{2} studies: \cite{Malaquias2017,Mackowiak2018} \\
        
        \midrule
        \multicolumn{3}{@{}l}{\textbf{\underline{Visual Metrics (\numberToBeChecked{6} studies)}}}        \\
        \multicolumn{2}{@{}l}{Eye tracking} & \numberToBeChecked{4} studies: \cite{Bauer2019,Binkley2013,Jbara2017,Fakhoury2019b} \\
        \multicolumn{2}{@{}l}{Letterboxing} & \numberToBeChecked{2} studies: \cite{Hofmeister2019,Schankin2018} \\
        
        \midrule
        \multicolumn{3}{@{}l}{\textbf{\underline{Brain Metrics (\numberToBeChecked{3} studies)}}}        \\
        \multicolumn{2}{@{}l}{fMRI} & \numberToBeChecked{1} study: \cite{Siegmund2017} \\
        \multicolumn{2}{@{}l}{fNIRS} & \numberToBeChecked{1} study: \cite{Fakhoury2019b} \\
        \multicolumn{2}{@{}l}{EEG} & \numberToBeChecked{1} study: \cite{Yeh2017} \\
        \bottomrule
    \end{tabular}
    }
\end{table}